\documentclass[conference,usenatbib]{basi}
\usepackage[T1]{fontenc}
\usepackage[british]{babel}
\usepackage[varg]{txfonts}
\usepackage{rotating}
\usepackage{dcolumn}
\begin{document}
\title[Nucleosynthesis inside accretion disks and outflows formed during core collapse of 
massive stars]{Nucleosynthesis inside accretion disks and outflows formed during core collapse of 
massive stars }
\author[I.~Banerjee]%
       {Indrani~Banerjee$^1$\thanks{email: \texttt{indrani@physics.iisc.ernet.in}}
       \\
       $^1$Department of Physics, Indian Institute of Science, Bangalore 560012, India}

\pubyear{2013}
\volume{**}
\pagerange{**--**}

\date{Received --- ; accepted ---}

\maketitle
\label{firstpage}

\begin{abstract}
We investigate nucleosynthesis inside the gamma-ray burst (GRB) accretion disks and in the outflows 
launched from these disks mainly in the context of Type~II collapsars. 
We report the synthesis of several unusual nuclei like 
$^{31}$P, $^{39}$K, $^{43}$Sc, 
$^{35}$Cl and various isotopes of titanium, vanadium, chromium, manganese and copper in the disk.
We also confirm the presence of iron-group and $\alpha-$elements in the disk, as shown by previous 
authors. 
Much of these 
heavy elements thus synthesized are ejected from the disk and survive in the outflows. 
While emission lines of several of these elements have been observed in the X-ray afterglows of GRBs
by BeppoSAX, Chandra, XMM-Newton etc., Swift seems to have not found these lines yet.
\end{abstract}


\section{Introduction}\label{s:intro}
We plan to investigate nucleosynthesis in 
the accretion disks formed by the Type II collapsars where the accretion rate ($\dot{M}$) 
is: $0.0001 M_{\odot} s^{-1} \lesssim \dot{M} \lesssim 0.01 M_{\odot} 
s^{-1}$, when $M_{\odot}$ indicates solar mass, as this regime is the ideal site for the synthesis 
of heavy elements. These disks are predominantly advection dominated. However,
neutrino cooling becomes important in the inner disk where the temperature and density are higher.
We also consider nucleosynthesis in the outflows from these disks and report that many of the heavy 
elements thus synthesized in the disk do survive in the outflow.
Moreover, depending on the abundance of $^{56}$Ni synthesized in the outflow, we can 
predict whether the outflow will lead to an observable supernova explosion or not. 

\begin{figure}
\centerline{\includegraphics[width=18cm]{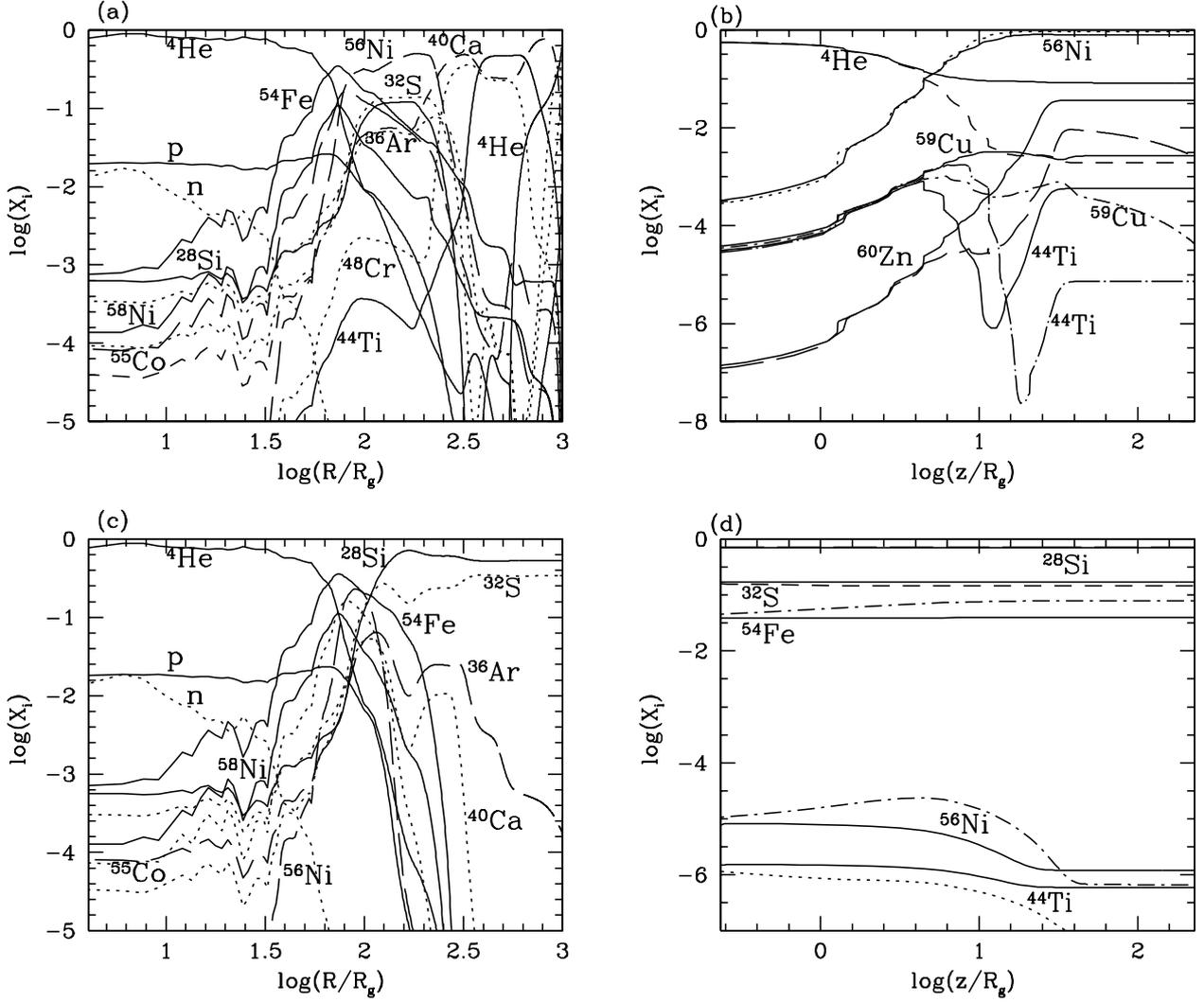}}
\caption{(a) \& (c) Zones characterized by dominant elements in the  
disk
with He-rich and Si-rich abundance at the outer disk respectively. (b) \& (d) Abundance 
evolution in the outflow from $R_{ej} \sim 40 R_g$ of the disk in (a), from $R_{ej} \sim 180 R_g$ of 
the disk in (c) respectively. In both (b) and (d)
solid lines correspond to the higher velocity of ejection and other lines to
lower velocity, in each set. }
\end{figure}

\section{Disk and Outflow Models}\label{s:fonts}
The accretion disk formed in a Type II collapsar is modelled within the 
framework suggested by 
\cite{Kohri} where the electron degeneracy pressure and the evolving neutron to proton 
ratio are appropriately calculated. Height-averaged equations based on a 
pseudo-Newtonian framework as suggested by \cite{Mukhopadhyay} is used .  
Following \cite{Fujimoto},  
we adopt a spherically
expanding, one-dimensional and adiabatic outflow model to investigate nucleosynthesis in the outflow.
Since $\dot{M}$ is 
very high, it is always possible that the matter may get deposited onto the 
accretion disk which favors outflow.
Outflows may also be due to magnetic centrifugal force and viscosity.
We use well tested nuclear network code as has been used by \cite{MC2000}.
We have modified this code  
further by increasing the nuclear network and including reaction rates from the JINA Reaclib Database,
https://groups.nscl.msu.edu/jina/reaclib/db/ \cite{Cybert}.
We use He-rich and Si-rich abundances as the initial conditions of nucleosynthesis at the outer disk.
We also consider outflow from various radii of ejection, $R_{ej}$, with $R_{ej} < 200 R_g $, 
$R_g$ being Schwarzschild radius
and evaluate
the abundance evolution in the outflow assuming the initial composition the same as in the accretion 
disk at $R_{ej}$. 

\section{Nucleosynthesis inside accretion disks and outflows}

Figures 1(a) and 1(c) illustrate the abundance evolution in the accretion disk
around a $3M_\odot$ Schwarzschild black hole accreting 
at  
$\dot{M}=0.001M_\odot s^{-1}$, with the viscosity parameter $\alpha=0.01$ and the composition of the 
accreting gas 
at the outer disk similar to the pre-supernova He-rich and Si-rich layer respectively. They depict that the 
disks
comprise of several zones characterized by dominant elements. In Fig. 1(a) the region $\sim 1000-300 
R_{g}$, is mainly the $^{40}$Ca, 
$^{44}$Ti and $^{48}$Cr rich 
zone. This is because unburnt $^{36}$Ar undergoes 
$\alpha$-capture reaction to give rise to $^{40}$Ca through $^{36}\rm Ar(\alpha,\gamma)^{40}Ca$, which 
undergoes partial $\alpha$-capture to give rise to $^{44}$Ti and $^{48}$Cr.
Inside this region, the temperature and density in the disk favor complete 
photodisintegration 
of $^{44}$Ti and $^{48}$Cr resulting in the formation of $^{40}$Ca,  
$^{36}$Ar, $^{32}$S and $^{28}$Si, as is evident from Fig. 1(a).
Subsequently, $^{28}$Si and $^{32}$S start burning, which favors formation of 
iron-group elements via {\it photodisintegration rearrangement} reactions \cite{Clayton}.
Therefore, in the range $\sim 300-80 R_{g}$, there is a zone overabundant 
in $^{56}$Ni, $^{54}$Fe, $^{32}$S and $^{28}$Si. 
Inside this zone, all the heavy elements photodisintegrate to $^{4}$He, neutron and proton.
In Figure 1(c) the disk has a huge zone rich in $^{28}$Si and $^{32}$S extending from $1000R_g$ to 
$250R_g$. 
Inside this radius, silicon burning commences and soon the disk becomes rich in $^{54}$Fe, $^{56}$Ni and
$^{58}$Ni. Inside $\sim 70 R_{g}$, all the
heavy elements again get photodisintegrated to $\alpha-$s and free nucleons. 
Another remarkable feature in the He-rich and Si-rich disks is that inside $\sim 100 R_{g}$, the 
abundances of various elements start becoming almost identical as if once threshold density and 
temperature are achieved, the nuclear reactions follow only the 
underlying disk hydrodynamics. \cite {BM} gives the details of the nuclear reactions. 
On increasing $\dot{M}$ ten times we find that the individual zones in both disks
shift outward retaining similar composition as is in the low $\dot{M}$ cases described above.

Next we consider outflow from $40 R_g$, which lies in the He-rich zone of the aforementioned He-rich disk.
The abundance evolution in the outflow is shown in Fig. 1(b). 
We find
that $^{56}$Ni is copiously synthesized along with isotopes of copper and zinc. Presence of 
$^{56}$Ni in the outflow 
signifies that it will result in an observable  
supernova 
explosion. Figure 1(b) also depicts that on changing the initial velocity of ejection the
final abundances of the nucleosynthesis products change significantly. More $^{56}$Ni is 
synthesized when the velocity of ejection is low (see Fig. 1(b)) because then the temperature drops slowly 
in the ejecta which facilitates greater recombination of alphas to nickel.
Figure 1(d) depicts the abundance evolution in the outflow from $180R_g$ of the above mentioned Si-rich 
disk. We choose this radius of ejection because outflow from the He-rich zone yields similar results as
in Fig. 1(b). Outflow from the Si-rich zone remains rich in $^{28}$Si and $^{32}$S. $^{56}$Ni is
hardly synthesized and there will be no observable supernova explosion.
\section{Summary and conclusions}\label{s:ADS}
Apart from the synthesis of iron-group and $\alpha-$elements
we report for the first time, to the best of our knowledge, that several unusual nuclei like $^{31}$P, 
$^{39}$K, $^{43}$Sc, $^{35}$Cl 
and various uncommon
isotopes of titanium, vanadium, chromium, manganese and copper are synthesized in the disk.
Several of these heavy elements survive in the outflow from these disks, and when $^{56}$Ni is
abundantly synthesized in the outflow, there is always a supernova explosion.

\section*{Acknowledgements}
 I would like to thank Banibrata Mukhopadhyay for suggesting the problem and discussing 
throughout the course of this work.
 This work was partly supported by the ISRO grant ISRO/RES/2/367/10-11.


\end{document}